\documentstyle[aps,prl,multicol]{revtex}
\begin{document}
\author{X. X. Yi,\ \ D. L. Zhou,  C. P. Sun}
\address{ Institute of Theoretical Physics, Academia Sinica, P.O.Box 2735, Beijing 100080, China}
\title{Decoherence in a single trapped ion due to  engineered reservoir}
\maketitle
\begin{abstract}
The decoherence in trapped ion induced by coupling the ion to the
engineered reservoir is studied in this paper. The engineered
reservoir is simulated by random variations in the trap frequency,
and the trapped ion is treated as a two-level system driven by a
far off-resonant plane wave laser field. The dependence of the
decoherence rate on the amplitude of the superposition state is
given.\\ {\bf PACS number(s):42.50.Vk, 03.65.-w,03.67.Lx}
\end{abstract}
\vspace{4mm}
\begin{multicols}{2}[]
\narrowtext

According to quantum mechanics[1], a system can exist in a
superposition of distinct states, whereas these superposition
state seems not to appear in the macroscopic world. One possible
explanation of this paradox[2] is based on the fact that systems
are never completely isolated but interact with the surrounding
environment, which contains a large number of degrees of freedom.
The environment influences the system evolution which continuously
decoheres and transforms system superposition into statistical
mixtures which behave classically[2,3]. There are many assumptions
involved in modeling the coupling of the system to its
environment. For example, the nature of the coupling of a system
to its environment is generally taken to be a linear[4] or a
nonlinear[5] function of the position operator of the object.
Assumptions are also made about the environment. One example is to
treat the environment as a reservoir of quantum oscillators, each
of which interacts with the quantum system in question. Such a
environment is extremely difficult to control because of the lack
of knowledge about the environment and its coupling to a system. A
recent work[6] extends the investigations of the decoherence
beyond the ambient reservoirs and engineer the state of the
reservoir, as well as the form of the system-reservoir coupling.
In the work, they apply noisy potentials to the trap electrodes to
simulate a hot reservoir, the system of trapped ion then feels a
controllable reservoir, through which quantum superpositions are
decohered into a state behaves classically.

In this letter, we present a theoretical study on decoherence of
quantum superpositions of a single trapped ion through coupling to
the engineered reservoir which is simulated by variating the trap
frequency. We tackle  the variations by treating them  as a white
noise, which results in decoherence of the trapped ion as you will
see further . Before introducing a engineered reservoir, we
consider a single $^9Be^+$ ion confined in an rf(paul) trap and
interacting with a plane wave laser field. The Hamiltonian of this
system may be written as (with $\hbar=1$)
\begin{eqnarray}
H&=&\frac{p^2}{2m}+V(r)+\frac{\omega_{eg}}{2}(|e\rangle\langle e|-|g\rangle\langle g|)\nonumber\\
&+&\frac{\Omega_L}{2}e^{i\omega_Lt-ik_Lr}|g\rangle\langle e|+H.c,
\end{eqnarray}
where the first three terms describe the free motion of the ion with two levels $|e\rangle$ and
$|g\rangle$ in a trap $V(r)=\frac 1 2 m \omega^2 r^2$, whereas the last two terms denote the coupling of
the ion to the plane wave laser field. Based on this kind of models, nonclassical motional states
such as thermal, Fock, coherent, squeezed and Schr\"odinger cat state are
created[7,8], it opens novel possibilities to study decoherence of quantum superposition[6].
For the plane wave excitation along the x-axis, the motional effects along the y and z directions
are unperturbed, in this case our Hamiltonian eq.(1) can simplifies to a one dimensional model
\begin{eqnarray}
\tilde{H}&=&\frac{p_x^2}{2m}+V(x)+\frac{\delta}{2}(|e\rangle\langle e|-|g\rangle\langle g|)\nonumber\\
&+&\frac{\Omega_L}{2}(|e\rangle\langle g|+|g\rangle\langle e|)+
p_xk_{Lx}(|e\rangle\langle e|-|g\rangle\langle g|),
\end{eqnarray}
where a unitary transformation
\begin{equation}
U_1=e^{-\frac i 2(\omega_Lt-k_{Lx}x)}|g\rangle\langle g|+e^{\frac i 2(\omega_Lt-k_{Lx}x)}|e\rangle\langle e|
\end{equation}
has also been made. The detuning $\delta=\omega_{eg}-\omega_L$,
the recoil shift $\frac{k_{L}^2}{2}$ is ignored here. The last
term in eq.(2) results from Doppler effect, which leads to the
coupling between the nearest neighbor motional states[9] and the
decoherence of a single trapped atomic/ionic qubit[10]. In the
following, we are concentrating on the situation where the
detuning $\delta$ and the coupling strength $\Omega_L$ is large,
especially, the coupling strength $\Omega_L$ is much larger than
the coupling of the system to ambient reservoir, the latter
condition is relevant to the recent engineered environment
experiment for avoiding the ambient environment which leads to the
natural decoherence[6]. Although the coupling length($\sim KHz$)
is large, it is smaller than the trap frequency($\sim MHz$) so
that the internal spin dynamics and the external motion of the ion
in the trap occur on two different time scales. Therefore, it is
useful to go to a rotating frame that eliminates the fourth term
in eq.(2), which describes the Rabi oscillations between the two
internal states. In the rotating frame, we will be able to
understand more clearly how the last term in eq.(2), which couples
the motional and the internal dynamics, leads to decoherence when
the trap frequency $\omega$ is modulated to be random.

We go to the rotating frame by making a unitary transformation
\begin{equation}
U_2=e^{-i(\frac{\Omega_L}{2}\sigma_x+\frac{\delta}{2}\sigma_z)t},
\end{equation}
where $\sigma_x=|g\rangle\langle e|+|e\rangle\langle g|$, $\sigma_z=|e\rangle\langle e|-
|g\rangle\langle g|$ are pauli matrices.
In the rotating frame, the system evolves according to
\begin{equation}
i\frac{\partial}{\partial t}|\psi^I(t)\rangle=H_I(t)|\psi^I(t)\rangle,
\end{equation}
where $H_I(t)$ is given by
\begin{eqnarray}
H_I(t)&=&\frac{p_x^2}{2m}+V(x)\nonumber\\
&+&p_xk_{Lx}(\alpha_x(t)\sigma_x+
\alpha_y(t)\sigma_y+\alpha_z(t)\sigma_z).
\end{eqnarray}
The time varying coefficients $\alpha_x(t),
\alpha_y(t),\alpha_z(t)$ are given by[11]
$$\alpha_x(t)=\frac{\Omega_L\delta}{\Omega_e^2}(1-cos(\Omega_et)),$$
$$\alpha_y(t)=\frac{\Omega_l}{\Omega_e}sin(\Omega_et),$$
$$\alpha_z(t)=\frac{\delta^2}{\Omega_e^2}+\frac{\Omega_L^2}{\Omega_e^2}cos(\Omega_et),$$
where $\Omega_e=\sqrt{\Omega_L^2+\delta^2}$. In what follows, we
make two simplifications in order to extract out the dominant
behaviors of the system. Noticing the coefficients $\alpha_i(t),
i=x,y,z$ oscillate rapidly, we expect the system in the rotating
frame  to evolve on a much slower time scale than the period
$2\pi/\Omega_e$. In this sense we can simplify the Hamiltonian
$H_I(t)$ given in eq.(6) by taking a average values of the
coefficients $\alpha_i(t), i=x,y,z$, this is equivalent to coarse
graining eq.(6). The coefficients $\alpha_i(t)$ then become time
independent and reduced to
$\alpha_x=\frac{\delta\Omega_L}{\Omega_e^2},$ $\alpha_y=0$,
$\alpha_z=\frac{\delta^2}{\Omega_e^2}$. Furthermore, we make
assumption that the system is being driven far-off-resonance, i.e,
$\delta >>\Omega_L $. We therefore have
\begin{equation}
H_I=\frac{p_x^2}{2m}+V(x)+p_xk_{Lx}\frac{\delta^2}{\Omega_e^2}\sigma_z,
\end{equation}
we note $p_x=i\sqrt{\frac{m\omega}{2}}(a^{\dagger}-a)$ with $ a(a^{\dagger})$ the annihilation (creation)
operator for motional state $|n\rangle$ which satisfies
$(\frac{p_x^2}{2m}+V(x))|n\rangle=\omega(n+\frac 1 2 )|n\rangle,$ therefore the Hamiltonian(7)
can be rewritten as
\begin{equation}
H_I=\omega a^{\dagger}a+ig\sigma_z(a^{\dagger}-a),
\end{equation}
where
$g=\sqrt{\frac{m\omega}{2}}\frac{\delta^2}{\Omega_e^2}k_{Lx}.$
$H_I$ couples nearest neighbors of motional state $|n\rangle$ with
a same internal level $|g\rangle$ or $|e\rangle$. As will be clear
further, the second term in $H_I$ leads to decoherence while the
engineered reservoir applied. To show this, we first of all give
the time evolution operator in
 the rotating frame,
\begin{eqnarray}
U_I(t)&=&e^{-i\omega a^{\dagger}at}(|g\rangle\langle g|e^{f(t)}e^{A(t)a^{\dagger}}e^{B(t)a}\nonumber\\
&+&|e\rangle\langle e|e^{f(t)}e^{-A(t)a^{\dagger}}e^{-B(t)a}),
\end{eqnarray}
where
$$A(t)=\frac{ig}{\omega}(e^{i\omega t}-1), B(t)=-A^*(t),$$
$$f(t)=-i\frac{g^2}{\omega}t+\frac{g^2}{\omega^2}(1-e^{-i\omega t}).$$
We consider a initial state of the form in the rotating frame
\begin{equation}
|\psi^I(0)\rangle=c_g|g\rangle\otimes|\alpha_g\rangle+c_e|e\rangle\otimes|\alpha_e\rangle,
\end{equation}
where $|\alpha_i\rangle, (i=g,e)$ denotes a coherent state, $c_g$
and $c_e$ are constants and satisfying $|c_g|^2+|c_e|^2=1$. This
kind of states may be created by during Raman transitions[7].
Eqs(3),(4) and (9) together govern the evolution of the system.
With these equations, we can analytically evolve the initial state
(10) to obtain
\begin{eqnarray}
|\psi(t)\rangle &=& e^{\frac i 2 \omega_L t-i\omega a^{\dagger}at}(\alpha_1(t)c_g(t)|\alpha_g^-(t)\rangle
\nonumber\\
&+&
\alpha_2(t)c_e(t)|\alpha_e^-(t)\rangle)\otimes|g\rangle\nonumber\\
&+&e^{-\frac i 2 \omega_Lt-i\omega a^{\dagger}at}(\alpha_2(t)c_g(t)|\alpha_g^+(t)\rangle\nonumber\\
&+&\alpha^*_1(t) c_e(t)|\alpha_e^+(t)
\rangle)\otimes|e\rangle,
\end{eqnarray}
where
$$\alpha_1(t)=cos\frac{\Omega_e}{2}t-\frac{i\delta}{\Omega_e}sin\frac{\Omega_e}{2}t,
\alpha_2(t)=-i\frac{\Omega_L}{\Omega_e}sin\frac{\Omega_e}{2}t,$$
$$c_g(t)=c_ge^{f(t)-\frac 1 2 |A(t)|^2},$$
$$c_e(t)=c_ee^{f(t)-\frac 1 2 |A(t)|^2},$$
$$\alpha_g^{\pm}(t)=\alpha_g+A(t)\pm \frac i 2
k_{Lx}\sqrt{\frac{1}{2m\omega}}e^{-i\omega t},$$
$$\alpha_e^{\pm}(t)=\alpha_e-A(t)\pm \frac i 2
k_{Lx}\sqrt{\frac{1}{2m\omega}}e^{-i\omega t}.$$ Eq.(11) is the
main result of our study, with which we can explain the essential
properties of the system. As eq.(11) shows, the initial motional
state  $|\alpha_g\rangle$ and $|\alpha_e\rangle$ are displaced to
be $|\alpha_g^{\pm}(t)\rangle$ and $|\alpha_e^{\pm}(t)\rangle$,
respectively. Their displacements depend on the internal state of
trapped ion. We are now interested in the coefficient of the
off-diagonal element of the density operator
$\rho(t)=|\psi(t)\rangle\langle \psi(t)|$ in ionic internal space,
its module which represents qualitatively decoherence of the
system is
\begin{eqnarray}
R(\omega,t)&=& Mod\{ \alpha_1^*(t)\alpha_2(t)|c_g|^2\langle\alpha_g^-(t)|\alpha_g^+(t)\rangle
e^{-i\omega \alpha_g^{+}(t)\alpha_g^{-*}(t)t}\nonumber\\
&+&\alpha_1^{*2}(t)c_g^*c_e\langle\alpha_g^-(t)|\alpha_e^+(t)\rangle
e^{-i\omega \alpha_e^{+}(t)\alpha_g^{-*}(t)t}\nonumber\\
&+&|\alpha_2(t)|^2c_e^*c_g\langle\alpha_e^-(t)|\alpha_g^+(t)\rangle
e^{-i\omega \alpha_e^{-}(t)\alpha_g^{+*}(t)t}\nonumber\\
&+&\alpha_1^*(t)\alpha_2^*(t)|c_e|^2\langle\alpha_e^-(t)|\alpha_e^+(t)\rangle
\nonumber\\
&\dot & e^{-i\omega \alpha_e^{-}(t)\alpha_e^{+*}(t)t} \} e^{-4\frac{g^2}
{\omega^2}(1-cos\omega t)}.
\end{eqnarray}
Where $Mod\{...\}$ denotes the module of the term in the brace.
We take a special case $\alpha_g=\alpha_e=2$
with fixed $\omega=2\pi\times 11.3 MHz$ as an example to illustrate
$R(\omega,t)$ vs. time in figure 1. As figure 1 shows,
 $R(\omega,t)$ is a periodic function of time $t$, which period depends on
the parameters chosen. During the first few Rabi cycles, the
coherent state parameters $\alpha_g^{\pm}(t)\simeq \alpha_g$
$\alpha_e^{\pm}(t)\simeq \alpha_e$, so that the internal and
external degrees of freedom appear to be decoupled and the system
simply oscillates rapidly between internal states, this was shown
in the first envelope of figure 1. However, for a long time
scales, the coupling between the internal and external states make
effects, this result in a modulation of the Rabi oscillations. Up
to now, the engineered reservoir does not take place and the
results illustrated in figure 1 indicate that without engineered
reservoir there are not any decoherences occur in the system.

A engineered reservoir coupled to the trapped ion is simulated by
variations in the trap frequency, oscillating near the ion's
original trapped frequency. Physically, decoherence in this case
arises from random perturbations of the Hamiltonian. In what
follows, we want to model the effects of the variations in the
trap frequency with the same formalism in ref.[12]. Our main idea
is to treat the variations in the trap frequency as fluctuations.
For an ion in a Paul trap[6], Colorado group realizes the
variations by a random voltage noise source applied to the trap
electrodes, the noise source is passed through a low-pass filter
network with a cut-off frequency well below the trap frequency.
The atom then sees a harmonic potential with fluctuating spring
constant. The Hamiltonian for a trapped ion in a harmonic
potential with fluctuating spring constant in the rotating frame
is
\begin{equation}
H_I=\frac{p_x^2}{2m}+V(x)+p_xk_{Lx}\frac{\delta^2}{\Omega_e^2}\sigma_z
+\frac 1 2 m\omega^2\varepsilon(t)x^2,
\end{equation}
this Hamiltonian is just the eq.(7) plus a term which describes the fluctuations in the
trap frequency.
If we take the fluctuations as a white noise, i.e.,
\begin{equation}
\varepsilon(t) dt=\sqrt{\Gamma} dW(t),
\end{equation}
and set $$X=\sqrt{\frac{m\omega}{2}}x,\ \ P_x=(2m\omega)^{-\frac 1
2 }p_x,$$ the Hamiltonian (14) becomes
\begin{eqnarray}
H&=&\omega(P_x^2+X^2)+gP_x\sigma_z+\sqrt{\Gamma}\omega X^2
dW(t)\nonumber\\
&=&H_0+\sqrt{\Gamma}\omega X^2 dW(t).
\end{eqnarray}
Here, $dW(t)$ is the increment of a real Wiener process[13],
$g=\sqrt{2m\omega}k_{Lx}\frac{\delta^2} {\Omega_e^2}$ and $\Gamma$
scales the fluctuations. For a single run with a known behavior of
the fluctuations in time, we use a stochastic Schr\"odinger
equation in the Ito formalism[14]
\begin{equation}
d\rho_I(t)=-\frac{i}{\hbar}[H_0,\rho_I]-\frac{i\sqrt{\Gamma}}{\hbar}
[X^2,\rho_I]dW(t)-\frac{\Gamma}{2}[X^2[X^2,\rho_I]]
\end{equation}
to describe the time evolution of the density operator $\rho_I$ in
the rotating frame. Here, we are not interested in the effects of
the fluctuation in short time scale, in this sense, we may take a
average over the fluctuation to get the master equation for the
average density operator $\rho^a$
\begin{equation}
\frac{d\rho^a}{dt}=-i[H_0,\rho^a]-\Gamma[X^2[X^2,\rho^a]].
\end{equation}
We want to determine the off-diagonal element of the density
operator $\rho=U_2\rho^aU_2^{\dagger}$ in the internal space
spanned by $|e\rangle$ and $|g\rangle$, where $U_2$ is given by
eq.(4). To do this, we first derive a system of equations for
$\rho^a_{ij}=\langle i|\rho^a|j\rangle, (i,j=g,e)$
\begin{eqnarray}
\frac{d\rho^a_{ge}}{dt}&=&2igP_x\rho_{ge}-\frac{\Gamma^2}{2}\omega^2[X^2[X^2,\rho^a_{ge}],\nonumber\\
\frac{d\rho^a_{eg}}{dt}&=&-2igP_x\rho_{eg}-\frac{\Gamma^2}{2}\omega^2[X^2[X^2,\rho^a_{eg}],\nonumber\\
\frac{d\rho^a_{ii}}{dt}&=&-\frac{\Gamma^2}{2}\omega^2[X^2[X^2,\rho^a_{ii}],(i=e,g)
\end{eqnarray}
It is easy to show that the off-diagonal element of the density operator $\rho$ can
be represented as
\begin{eqnarray}
\langle g|\rho|e\rangle &=& \alpha_1^*(t)\alpha_2^*(t)\rho^a_{gg}(t)+
[\alpha_1^*(t)]^2\rho^a_{ge}(t)\nonumber\\
&+&|\alpha_2(t)|^2\rho^a_{eg}(t)+\alpha_1^*(t)\alpha_2(t)\rho^a_{ee}(t).
\end{eqnarray}
For a short time scale, $\langle n|\rho_{ij}^a(t) |m\rangle\sim
0$, for $m\neq n$,  The module $R(t)$ of the off-diagonal element
which characterizes the decoherence is given by
\begin{eqnarray}
R(t)&=&Mod\{ Tr_e\langle g|\rho|e\rangle\} \\ &=&Mod \{
\sum_{n=0}^{\infty} [\alpha_1^*(t)\alpha_2^*(t)\langle
n|\rho^a_{gg}(0)|n\rangle\nonumber\\ &+& [\alpha_1^*(t)]^2\langle
n|\rho^a_{ge}(0)|n\rangle\nonumber\\ &+&|\alpha_2(t)|^2\langle n|
\rho^a_{eg}(0)|n \rangle\nonumber\\
&+&\alpha_1^*(t)\alpha_2(t)\langle n| \rho^a_{ee}(0)|n\rangle ]
e^{-\Gamma\omega^2(n^2+n+1)t}\} ,
\end{eqnarray}
where $\alpha_1(t)$, $\alpha_2(t)$ and $|n\rangle$ are the same as
above mentioned, $Tr_e$ denotes a trace over the external states.
As eq.(21) shows, the decoherence rate $\Gamma\omega^2(n^2+n+1)$
depends on the character of the fluctuations, the trap frequency
and the motional state of the trapped ion. Physically, the trap
frequency play a role of the coupling of the system to the
reservoir, so the larger the trap frequency, the larger the
decoherence rate(decay rate). The fact that the decoherence rate
depends on the motional state of trapped ion was observed in the
experiments of ref.[7]. Again, we consider the state given by
eq.(10) as a initial state, the numerical results of eq.(20) are
illustrated in Fig.2, there are a few Rabi cycles in the
beginning. However, for a long time scales, the oscillations
disappear, it is evidence that decoherence occur in the system.

To sum up, decoherence in a two-level trapped ion is studied in
this paper. The decoherence is induced by coupling the ion to the
engineered reservoir, which is simulated by random variations in
the trap frequencies. Without this reservoir, the transitions
between the ionic internal levels manifest modulated Rabi
transition, whereas the transition was suppressed when the
engineered reservoir take place. The suppressed transitions
indicate that there is decoherence in the trapped ion system.\\
{\bf  ACKNOWLEDGEMENT:} This work was supported by Chinese
postdoctoral Fund via Institute of Theoretical Physics, Academia
Sinica. Discussions with Prof. W.M.Zheng, Dr. S. X. Yu and Dr.
Y.X. Liu are gratefully acknowledged.\\

\ \ \\ {\bf Figure captions:}\\ {\bf Fig.1:} The module of the
off-diagonal element of the density operator is plotted as a
function of time with fixed trap frequency. The parameters chosen
are $\alpha_g=\alpha_e=3$, $\omega=2\pi\times 11.3 MHz$,
$\delta=4.0GHz$, $\Omega_L=10.0KHz$.\\ \ \ \\
{\bf Fig.2:} This
plot shows the decoherence in trapped ion induced by the
engineered reservoir. The module of the off-diagonal element of
the density operator is calculated by the master equation. This
plot shows the module as a function of time. The parameters chosen
are the same as in figure 1 and $\Gamma=1.0KHz$.
\end{multicols}

\begin{thebibliography}{99}
\bibitem.L.D.Landau and E.M.Lifshitz, Quantum Mechanics (Pergamon press,1977) Third revised edition.
\bibitem.W.H.Zurek, Phys. Today 44, No.10 (1991)36.\\
W.H.Zurek, Phys. Rev. D 24 (1981) 1516, {\it ibid}26 (1982)
1862.\\ C.P.Sun, X.X.Yi and X.J.Liu, Fortschr. Phys. 43 (1995)
585.
\bibitem. D.F.Walls, G.J.Milburn, Phys. Rev. A 31(1985) 2403.\\
A.O.Caldeira, A.J.Leggett, Phys. Rev. A 31(1985)1509.
\bibitem.A.O.Calderia, A.J.Leggett, Physica A 121(1983)587.\\
E.Joos, H.D.Zeh, Z. Phys. B 59(1985)223.\\
W.G.Unruh, W.H.Zurek, Phys. Rev. D 40(1989)1071.\\
J.P.Paz, S.Habib, W.H.Zurek, Phys. Rev. D 47(1993)488.
\bibitem.B. L. Hu, J. P. Paz, Y. H. Zhang, Phys. Rev. D 47 (1993) 1576.\\
L.M.Kuang, H.S.Zeng, Z.Y.Tong, Phys. Rev. A 60(1999)3815.
\bibitem. C.J.Myatt, B.E.King, Q.A.Turchette, C.A.Sackett, D.Kielpinski,
W.M.Itano, C.Monroe, D.J.Wineland, Nature 403(2000)1269.
\bibitem.D. M. Meekhof, C. Monroe, B. E. King, W. M. Itano, D. J. Wineland, Phys. Rev. Lett. 76(1996)1796.\\
C.Monroe, D.M.Meekhof, B.E.King, D.J.Wineland, Science 272(1996)1131.
\bibitem.R.L.de Matos Filho, W.Vogel Phys. Rev. Lett. 76(1996)608.
\bibitem.J.Javanainen {\it etal.}, J.Opt.Soc. Am. B 1(1984)111.
\bibitem. L. You, Quant-ph/0001117.
\bibitem. J.Williams, R.Walser, J.Cooper, E.A.Cornell, M.Holland,
Phys. Rev. A 61(2000)033612.
\bibitem. S.Schneider, G.J.Milburn, Phys. Rev. A 57 (1998) 3748.\\
{\it ibid,} 59(1999) 3766.
\bibitem. C.W. Gardiner, Handbook of stochastic process for physics, Chemistry and the natural science
(Springer-Verlag, Berlin,1985).
\bibitem. S.Dyrting, G.J.Milburn, Quantum and Semiclass. Opt. 8(1996)541.

\end{thebibliography}
\end{document}